\documentclass[fleqn,10pt]{wlscirep}
\usepackage[utf8]{inputenc}
\usepackage[T1]{fontenc}
\usepackage{lineno}
\usepackage{amsmath}
\usepackage{subcaption}
\usepackage{graphicx}

\usepackage[format=plain,justification=justified,singlelinecheck=false,font={stretch=1.125,small,sf},labelfont=bf,labelsep=space]{caption}

\newcommand{\ie}[0]{\textit{i.e.}, }
\newcommand{\aka}[0]{\textit{a.k.a.}, }
\newcommand{\eg}[0]{\textit{e.g.}, }

\newcommand{\via}[0]{\textit{via} }
\newcommand\norm[1]{\left\lVert#1\right\rVert}

\usepackage{chemformula} 
\usepackage{braket}
\newcommand{\DefineAuthor}[2]{%
  \expandafter\newcommand\csname #1note\endcsname[1]{%
    \textbf{\textcolor{#2}{\textbf{#1:} ##1}}}%
  \expandafter\newcommand\csname #1\endcsname[1]{
    \textbf{\textcolor{#2}{##1}}}
  \expandafter\newcommand\csname #1cancel\endcsname[1]{%
    \textbf{\textcolor{#2}{\sout{##1}}}}%
  \expandafter\newcommand\csname #1change\endcsname[2]{%
    \textbf{\textcolor{#2}{\sout{##1} ##2}}}%
  \newenvironment{#1text}{\color{#2}}{\color{black}}
}
\definecolor{dartmouthgreen}{rgb}{0.05, 0.5, 0.06}
\DefineAuthor{LMS}{blue}
\DefineAuthor{AF}{brown}

\title{
Enabling Inverse Design in Chemical Compound Space: Mapping Quantum Properties to Structures for Small Organic Molecules
}

\author[1]{Alessio Fallani}
\author[1]{Leonardo Medrano Sandonas}
\author[1]{Alexandre Tkatchenko}
\affil[1]{Department of Physics and Materials Science, University of Luxembourg, L-1511 Luxembourg City, Luxembourg.}
\affil[*]{Corresponding authors:  Alessio Fallani (alessio.fallani@uni.lu), Leonardo Medrano Sandonas (leonardo.medrano@uni.lu), Alexandre Tkatchenko (alexandre.tkatchenko@uni.lu)}

\begin{abstract} 
Computer-driven molecular design combines the principles of chemistry, physics, and artificial intelligence to identify novel chemical compounds and materials with desired properties for a specific application.
In particular, quantum-mechanical (QM) methods combined with machine learning (ML) techniques have accelerated the estimation of accurate molecular properties, providing a direct mapping from 3D molecular structures to their properties.
However, the development of reliable and efficient methodologies to enable \emph{inverse mapping} in chemical space is a long-standing challenge that has not been accomplished yet.
Here, we address this challenge by demonstrating the possibility of parametrizing a given chemical space with a finite set of extensive and intensive QM properties.
In doing so, we develop a proof-of-concept implementation that combines a Variational Auto-Encoder (VAE) trained on molecular structures with a property encoder designed to learn the latent representation from a set of QM properties.
The result of this joint architecture is a common latent space representation for both structures and properties, which enables  property-to-structure mapping for small drug-like molecules contained in the QM7-X dataset.
We illustrate the capabilities of our approach by conditional generation of \emph{de novo} molecular structures with targeted properties, transition path interpolation for chemical reactions as well as insights into property-structure relationships. 
Our findings thus provide a proof-of-principle demonstration aiming to enable the inverse property-to-structure design in diverse chemical spaces.

\end{abstract}
\begin{document}

\flushbottom
\maketitle

\thispagestyle{empty}

\section{Introduction} 

The discovery and optimization of chemical compounds can be accelerated thanks to the marked advancements in quantum and statistical methods, their implementation in advanced software, as well as the seemingly never-ending improvement in computer hardware.~\cite{Kulik22,Sadybekov23}
Unlike the  traditional painstaking trial-and-error process which heavily relied on experimental work (known as the Edisonian approach), we can now compute a wide range of accurate physicochemical properties of a given compound using quantum-mechanical (QM) methods with only the respective atomic coordinates and atom types.
However, rationally exploring the incredibly vast chemical compound space (CCS, estimated to contain $10^{60}$ molecular structures even for small organic molecules)~\cite{lilien} \via highly accurate QM methods is still unfeasible due to their prohibitive computational cost. 
In this regard, machine learning (ML) techniques have revolutionized the field of molecular design by offering a fast but just as accurate method for obtaining properties from 3D molecular structures (\aka direct mapping),~\cite{SchNet, ANI, wignerkernel, Batzner2022,Steinmann23} positioning them as an indispensable resource in high-throughput screening pipelines.~\cite{AIdrivenhts,BiasedHTS,datadrivenHTS,historicalHTS, efficientvirtualapache, publicdomainhts,librarysubsetscreening}  
Although these approximate mappings have undoubtedly enhanced our understanding of the CCS, it is the possibility to invert them that has the potential to truly disrupt and transform the field.
Addressing this challenge would allow us to predict the 3D molecular structures from their inherent properties, representing a paradigm shift in the design and discovery of chemical compounds with specific functionalities.

The quest to establish an inverse mapping has emerged as a formidable challenge, captivating the interest and dedication of researchers from various disciplines such as organic chemistry, materials science, and molecular docking.~\cite{Sanchez18,Zunger2018, Kim2018, Chen2023,Lee23,Moret2023, lin23, Noh20,Nigam21,Nigam22}
A hint for the existence of such property-to-structure relationship is given by considering the demonstrated capability of generative models to selectively generate random structures conditioned on a predefined set of desired properties.
Indeed, generative modelling has yielded numerous groundbreaking research outcomes,~\cite{Isayev,Seo23,Dollar21} particularly in the field of cheminformatics, in which there is an extensive literature delving into diverse generative architectures and research focuses.~\cite{bombarelli, molgan, olive, conditionaldesigndeepgen,diff_dock,Organ,nevae,multiobjectivedenovograph,molcyclegan,MolFlow,junctiontreevae,graphite}
Models dealing with the intricate 3D structure of molecules have been recently developed, leading to promising results across different design tasks.~\cite{3dgenrew,gschnet,equivdiff,diffusioncrystal, diffusionpriorbridge, diffusiontargetaware, conformer_diff, diff_dock}
In the same breath, multi-objective optimization problems have been tackled using generative models and genetic algorithms, \eg the design of functional organic molecules for optoelectronics applications and of candidate structures for dielectric organic materials.~\cite{ParetoOligomer,Polymerdesign_opt, Yuan20, Westermayr2023}
These compelling examples underscore the relevance of tackling the inverse design problem, yet the potential of parametrizing the CCS using properties as coordinates remains unexplored. 
Overcoming this challenge would lay the groundwork for an alternative and multifaceted approach to understanding and manipulating the intricate relationship between the properties and structures of organic molecules.

\begin{figure}[t!]
\centering
        \includegraphics[width=0.95\textwidth]{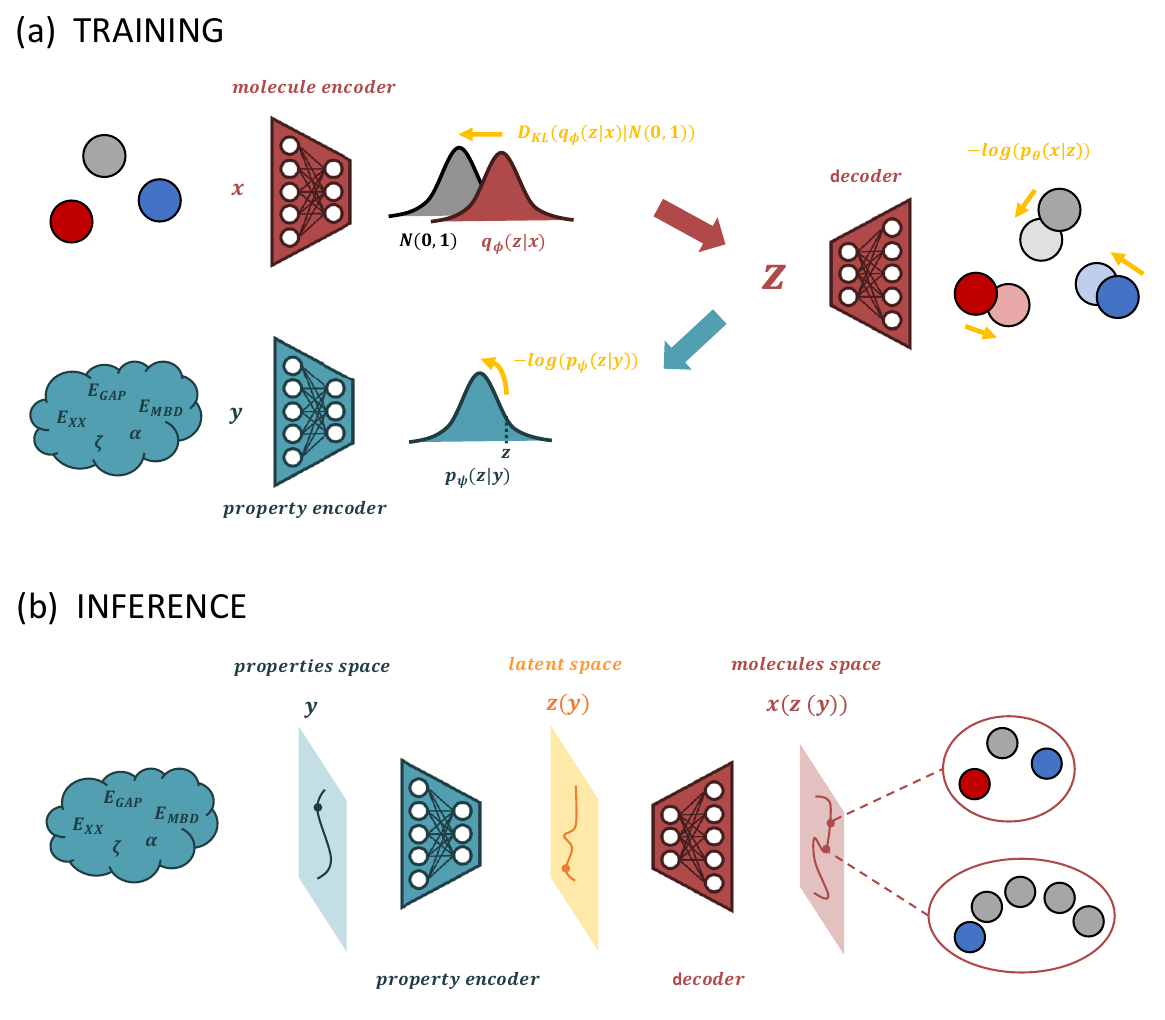}
        \caption{\textbf{Scheme of our approach to inverse property--structure design.}
        (a) For training our model, we used a Variational AutoEncoder (VAE) architecture to encode the molecular structures $x$ (represented as Coulomb matrices)  with a property encoder to encode the associated quantum-mechanical (QM) properties $y$.
        The loss function employed to train these models comes from a modified \emph{evidence lower bound} (ELBO), see Eq.~\ref{loss}.
        The result of this joint training is a common latent representation for both properties and structures. 
        (b) At inference time, one can combine the property encoder with the decoder component of the VAE and, hence, to successfully approximate the CCS parameterization using QM properties as intrinsic coordinates. 
        The differentiability of our CCS parameterization enables us to identify the most relevant  properties in the molecular reconstruction  process as well as to perform a series of molecular design tasks.} 
        \label{implementation}
\end{figure}

Stemming from the ``freedom of design'' conjecture in the molecular property space espoused in our previous works,~\cite{medranoquantum,Szabolcs23} this study aims to investigate the feasibility of learning the intricate parameterization of CCS by leveraging the comprehensive QM data of equilibrium small drug-like molecules contained in the QM7-X dataset.~\cite{qm7x}
In doing so, we propose a proof-of-concept implementation that combines a Variational Auto-Encoder (VAE) architecture to encode the molecular structures (represented as Coulomb matrices)  with a property encoder to encode the associated QM properties, see Fig.~\ref{implementation}. 
The joint training of the VAE and property encoders yields a low-dimensional internal representation that is common for both the molecular structures and QM properties. 
This enables us to combine the property encoder with the decoder component of the VAE and, hence, to successfully approximate the CCS parameterization using QM properties as intrinsic coordinates for navigating the QM7-X chemical space.
As a result, our method is able to accurately predict the heavy atom composition of molecules and to reconstruct their geometries with reasonably good accuracy. 
Moreover, thanks to the differentiability of our CCS parameterization, we are able to identify the most relevant  properties in the molecular reconstruction  process as well as substructures in the molecular property space covered by QM7-X. 
The conventional multi-objective generative modeling paradigm is also retrieved by conditionally sampling in the input space of properties to accomplish two distinct design tasks with well-defined objectives.
As a final showcase of the capabilities of learning a CCS parameterization, we have implemented a geodesic search algorithm which uses the latent space representation as internal coordinates, enabling the definition of transition structures and the estimation of rotational energy profiles by learning only from equilibrium geometries. 

Although our proof-of-concept implementation is currently assessed only for small molecules, our findings suggest that a CCS parameterization based on QM properties is feasible and holds promise for a wide range of applications, including interpretability, generation of \textit{de novo} molecules, transition path interpolation, and reaction barrier estimation.
Thus, this work highlights the remarkable opportunities that arise from defining an inverse mapping approach to advance our understanding of the chemical compound space by connecting QM properties and molecular structures.

\section{Background, implementation and methods} \label{sec:model}

\subsection*{Variational Auto-Encoder framework for inverse design}
A Variational Auto-Encoder (VAE) is in general comprised of two neural networks called \emph{encoder} and \emph{decoder}.
The \emph{encoder} encodes the input in a lower-dimensional latent space representation and the \emph{decoder} decodes that representation trying to recover the encoded input.
The model thus learns a continuous low-dimensional representation capturing the most salient statistical features of the input during the compression process. 
Whereas, in the decoding process, the decoder learns how to generate samples coherent with the ones in the dataset.
Both networks are probabilistic, meaning that they parameterize a probability distribution. 
This, together with a regularization term forcing the latent distribution to be close to a reference distribution, helps obtaining a less sparse internal representation and avoiding large portions of the latent space to decode into invalid outputs. 
The loss function used to train these models comes from the well known \emph{evidence lower bound} (ELBO).\cite{elbo} 
If $x$ is a sample from a dataset with probability distribution $p(x)$, then the loss reads:

\begin{equation}
    \mathrm{loss}_{\rm ELBO} = D_{\rm KL}[q_{\phi}(z|x)||p(z)] - \mathbb{E}_{q_{\phi}}[\log(p_{\theta}(x|z))] \; ,
\end{equation}

\noindent where $q_{\phi}(z|x)$ and $p_{\theta}(x|z)$ are the probability distributions parameterized  by the encoder and the decoder, respectively. 
$D_{\rm KL}$ is the Kullback-Leibler divergence and $p(z)$ is the prior probability distribution of the latent space representation $z$. 
Usually all distributions are chosen to be identically multivariate Gaussian distributed  and $p(z) = \mathcal{N}(0,\mathbb{I})$. 
In the training procedure, the \emph{encoder} learns to encode $x$ in the probability distribution $q_{\phi}(z|x)$ from which its latent representation $z$ is sampled using the well-known reparameterization trick.~\cite{elbo}
The divergence term here ensures that the representation stays compact, since the distribution will be as close as possible to the prior distribution $\mathcal{N}(0,\mathbb{I})$. 
Then, $z$ is fed to the \emph{decoder} which learns to turn this random variable back to $x$. 
After training is over, one can sample $z$ from $\mathcal{N}(0,\mathbb{I})$, feed it to the decoder and obtain samples $x$ with a distribution close to $p(x)$.

In our implementation, we start from a dataset $D = (X, Y)$ where $X$ are the molecular structures and $Y$ are their QM properties. 
To obtain an inverse mapping $f: Y \rightarrow X$, we modified the standard VAE framework by adding a third network parameterizing the probability distribution $p_{\psi}(z|y)$.
Moreover, we modify the loss function by including a likelihood term (\eg $-\log(p_{\psi}(z|y))$) in the ELBO loss function, which is now expressed as:

\begin{equation}\label{loss}
    \mathrm{loss} = \beta D_{KL}[q_{\phi}(z|x)||p(z)] - \mathbb{E}_{q_{\phi}}[\log(p_{\theta}(x|z))] - \tau\mathbb{E}_{q_{\phi}}[\log(p_{\psi}(z|y))] \; ,
\end{equation}

\noindent where $\beta$ and $\tau$ are adjustable coefficients introduced as hyperparameters.
%
%
The training procedure is here similar as described for VAE but, in this case, the variable $z$ sampled from $q_{\phi}(z|x)$ will also maximize the likelihood term $\mathbb{E}_{q_{\phi}}[\log(p_{\psi}(z|y))]$. 
This modification ensures the emergence of a common latent representation for both the VAE and the property \emph{encoders}. 
After training, one takes samples of $y$ from the property space and gets a value for $z$ as the mean of $p_{\psi}(z|y)$  that serves as the maximum-likelihood estimator for the latent space representation of the corresponding molecular structure. 
Subsequently, a molecule $x$ can be generated, which is expected to possess properties similar to $y$.
A schematic representation of our proposed implementation is shown in Fig.~\ref{implementation}.

\subsection*{Molecular representation and structure retrieval}
An ideal representation for inverse design must fulfil the following criteria: firstly, it has to encode the atomic positions and atomic species of a molecule with as many invariances as possible (translational, rotational and permutational). 
Secondly, it has to show a strong correlation with QM properties. 
Finally, and most importantly, it has to be invertible, namely that one should be able to retrieve the Cartesian coordinates and atomic species.
Neglecting the permutational invariance requirement, we find that a simple yet effective 3D molecular representation is the Coulomb matrix (CM).~\cite{Rupp,CMpred}
CM is an elegant and physically-inspired descriptor, which has shown great success in a wide variety of investigations related to molecular property prediction. 
CM is invariant to translations and rotations, allowing for the retrieval of atomic positions and species of a molecule up to a chirality transformation. 
This representation is defined as:

\begin{equation}
  M_{ij}=\begin{cases}
    0.5 Z_i^{2.4} \, & \text{if $i=j$}.\\
    \frac{Z_iZ_j}{|r_i-r_j|}, & \text{otherwise} \; .
  \end{cases}
\end{equation}

\noindent where the indices $i$ and $j$ run over the atoms in the molecule and $Z_i$ indicates the atomic species. 
We have chosen to treat the hydrogen (\ch{H}) atoms implicitly in our work due to several reasons. 
Primarily, the associated terms in the Coulomb matrix representation pertaining to H atoms are significantly smaller compared to the other atoms in a given system. 
As a consequence, even small errors in the reconstruction process can lead to disproportionately large changes in the positions of  H atoms. 
By treating the hydrogen atoms implicitly, we can mitigate the impact of these potential errors and minimize the distortion caused by inaccuracies in their positions. 
This approach allows us to focus our method on accurately representing the molecular scaffold (\ie positions of heavy atoms).
For what concerns the network input standardization, we adapted the matrix to the maximum possible number of atoms per species found in the dataset, padding the rest of the entries with zeros (see Sec. 1 of the Supplementary Information (SI)).

The retrieval of Cartesian coordinates and chemical composition from CM follows two steps.
We first obtain the composition from the diagonal elements by applying the inverse function $g=(2(\cdot))^{\frac{1}{2.4}}$ and rounding the outcomes to the closest integer values. 
Accordingly, we use the set $\{Z\}$ of atomic numbers obtained and get the distances $d_{ij}$ as $d_{ij} = \left(\frac{M_{ij}}{Z_iZ_j}\right)^{-1}$. 
Lastly, we apply \textit{classical multidimensional scaling} (CMDS) to the resulting Euclidian distance matrix (EDM)~\cite{edms} to get the Cartesian coordinates of atoms up to a chirality transformation (see Sec. 1 of SI). 
To reconstruct H atoms, we use OpenBabel software ~\cite{babel} that restores connectivity and bond order based on interatomic distances. 
After this, the positions of the H atoms are optimized with third-order self-consistent charge density-functional tight binding (DFTB3)~\cite{seifert96,gaus11} that also accounts for many-body dispersion/van der Waals (vdW) interactions \via the MBD approach~\cite{tkatchenko12,stoehr16} (DFTB3+MBD, same level of theory used for optimizing the geometries in QM7-X dataset), while freezing the position of the rest of the atoms in the molecule.

As a final remark concerning the choice of representation, a graph-based representation would be beneficial for generative purposes as it treats molecular fragments robustly and avoids the need for truncation.
However, in this work, we are aiming at defining a CCS parameterization based on a set of global QM properties of molecules. 
Consequently, it becomes crucial to account for the potential variation in the number of atoms as we navigate through the property coordinates.
Given the challenges associated with graph-based approaches due to the requirement of specifying the number of nodes in advance, the utilization of (appropriately padded) CM proves to be an efficient and scalable representation for inverse design purposes at this stage of model development.  
%

%
%
\renewcommand{\arraystretch}{1.2} 
\begin{table}[ht!]
  \centering
    \begin{tabular}{c l c c c}
    \hline\hline
    \textbf{Symbol} & \multicolumn{1}{l}{\textbf{Property Description}} & \textbf{Units}  &  \textbf{Type} & \textbf{Class} \\
    \hline
    $E_{AT}$ & Atomization energy & eV &   M,G  & E \\
    $E_{MBD}$ & MBD energy & eV &   M,G & E \\
    $E_{XX}$ & Exchange energy & eV &   M,G & E \\
    $E_{NN}$ & Nuclear-nuclear energy & eV &   M,G & E \\
    $E_{EE}$ & Electron-electron energy  & eV &   M,G & E \\
    $E_{KIN}$ & Kinetic energy & eV &   M,G & E \\
    $E_{GAP}$ & \textsc{homo-lumo} gap & eV &   M,G  & I \\
    $E_{HOMO}^{0}$ & \textsc{homo} energy & eV &    M,G  & I \\
    $E_{LUMO}^{0}$ & \textsc{lumo} energy & eV &   M,G  & I \\
    $E_{HOMO}^{1}$ & \textsc{homo}-1 energy & eV &    M,G  & I \\
    $E_{LUMO}^{1}$ & \textsc{lumo}+1 energy & eV &   M,G  & I \\
    $E_{HOMO}^{2}$ & \textsc{homo}-2 energy & eV &    M,G  & I \\
    $E_{LUMO}^{2}$ & \textsc{lumo}+2 energy & eV &   M,G  & I \\
    $\zeta$ & Total dipole moment & $e\cdot$\AA{} &    M,G   & I \\
    $\alpha$ & Isotropic molecular polarizability & $a_0^3$ &    M,R   & E \\
    $D_{MAX}$ & Maximum atom-atom distance  & \AA{} &   S,G  & I \\
    \hline\hline 
  \end{tabular}
  \caption{
  \textbf{Quantum mechanical (QM) properties.}
  List of QM properties (and corresponding symbols) taken from the QM7-X dataset~\cite{qm7x} and considered during the training of our model, see Fig.~\ref{implementation}.
  In the units provided for each of these QM properties, $a_0$ represents the atomic units of length (Bohr radius).
  All properties are scalars, \ie their dimension is 1.
  Property types and classes were categorized as follows: structural (S), global/molecular (M), ground-state (G), response (R), extensive (E), and intensive (I).
  }
  \label{table1}
\end{table}
%
%

\section{Results and Discussion} 

\subsection{Property-to-structure mapping}

To train the VAE and the property encoder (see Sec.~\ref{sec:model}), we have considered a subset from QM7-X dataset of $40,988$ equilibrium conformations of molecules with up to seven heavy atoms including C, N, and O. 
For training our model, we used the following data splitting: 28k and 2k molecules for the training and validation sets, respectively, while the remaining molecules were used for testing the model. 
Moreover, since the QM7-X dataset contains a plethora of physicochemical properties, computed  by  means of non-empirical hybrid density-functional theory (DFT) and a many-body treatment of vdW/dispersion interactions (\ie PBE0+MBD)~\cite{pbe0a,adamo1999,tkatchenko12,Ambrosetti2014} in conjunction with the tightly converged numeric atom-centered basis sets,~\cite{havu2009efficient} we have opted to consider 17 global (extensive and intensive) properties during the training process. 
The selected properties are listed in Table~\ref{table1}.
The technical details of the neural networks defined for the VAE and the property encoder are specified in Sec. 2 of the SI. 

\begin{figure}[t!]
\centering
        \includegraphics[width=0.9\linewidth]{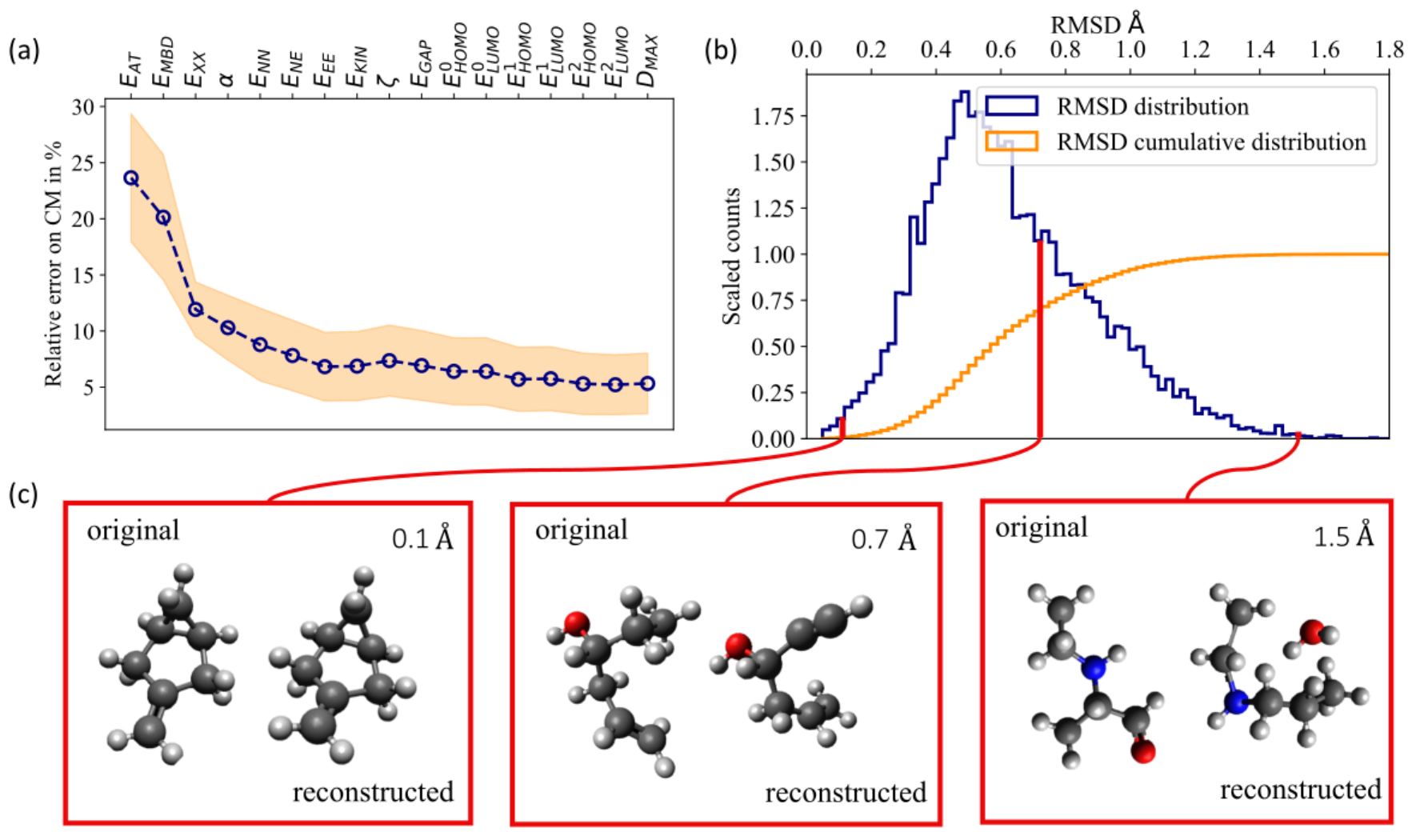}
    \caption{\textbf{Performance of the trained model in the molecular reconstruction process.} 
    (a) Average relative error on the reconstruction of Coulomb matrix (CM) representation of structures from QM properties as a function of the number of molecular properties considered during the training. The orange shadow represents the standard deviation per case.
    (b) Frequency plot of the root-mean squared error (RMSD) between the geometries retrieved from the reconstructed CMs and corresponding original representations. The cumulative function of this plot is also presented, see orange curve. 
    (c) Select examples for different RMSD values to compare the original molecule with  this one reconstructed from a predefined set of QM properties. We empirically found that RMSD $<0.7$ \r{A} is an adequate threshold to separate molecules with an acceptable reconstruction of the heavy atom structure in terms of topology and orientation.}
    \label{reconstructions}
\end{figure}

We now assess the capability of the trained model to establish an approximate parameterization of the chemical compound space (CCS) spanned by QM7-X based on a predefined set of QM properties.
In doing so, we will make use of the molecules in the test set together with their corresponding properties, \ie the property set of these molecules will be used to construct the model and, then, the generated molecule will be compared with the original one.
Fig.~\ref{reconstructions}(a) shows the average value of the relative error on Coulomb matrix (CM) reconstruction over the test set for an increasing number of properties used to train the model. 
We have analyzed the relative error on CM instead of the root-mean squared deviation (RMSD) between structures because the latter is only defined for molecules for which the composition is correctly predicted and, consequently, it will present more fluctuations across different number of properties especially when the number of properties used is low and hence the error is high.
Accordingly, one can see that our model allows to reconstruct the representation with an average error that converges to $\sim 5\%$ when considering more than seven properties during the training. 
This relative error is defined as $\Delta = \frac{|\tilde{\mathbf{C}} - \mathbf{C}|}{|\mathbf{C}|}\times 100$, where $\mathbf{C}$ is the original CM and $\tilde{\mathbf{C}}$ is the reconstructed one (both to be considered vectors). 
While we provide the mean and deviation of the CM reconstruction error, it is important to note that this indicator does not completely reflect the quality of the mapping due to its noisy and nonlinear correlation with RMSD (see Fig. S1 of SI).
To have a better understanding of this, we also report the distribution and the cumulative distribution over the test set for the RMSD in the case of full set of properties (see Fig.~\ref{reconstructions}(b)). 
%
%
Indeed, the mode of the RMSD distribution is close to $0.5$ \r{A} and $\approx 75\%$ of the test set is reconstructed with an RMSD $<0.7$ \r{A}.
Despite obtaining a wide RMSD spectrum ([0.05, 1.6] \AA{}),  we found that $\sim 70\%$ of the molecules in the test set are reconstructed within RMSD $=0.7$ \r{A}. 
We empirically found this threshold to be adequate to separate molecules with an acceptable reconstruction of the heavy atom structure in terms of topology and orientation (see illustrative examples of original and reconstructed molecules in Fig.~\ref{reconstructions}(c)). 
Concerning the chemical composition reconstruction, the model exhibits excellent performance, correctly predicting it for $99.96\%$ of the molecules in the test set.
A detailed explanation elucidating these results will be discussed in the following section.

\begin{figure}[t!]
    \centering
    \includegraphics[width=0.6\linewidth]{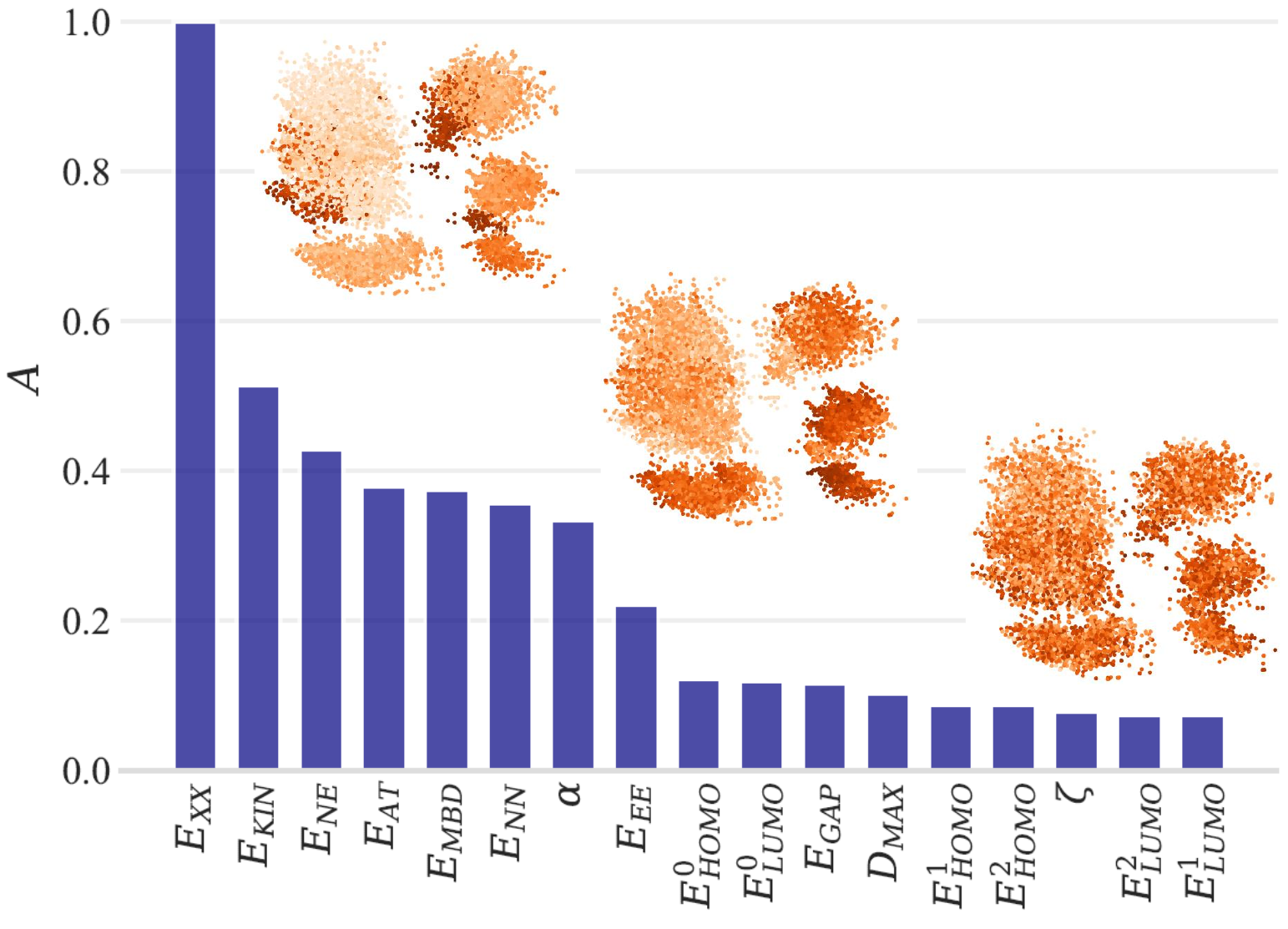}
    \caption{\textbf{Gradient attribution map for molecular properties.} 
    The contribution of each property $p_j$ to the output structure is evaluated by defining the variable $A$ which considers the partial derivative of the reconstructed CM with respect to $p$, see Eq.~\ref{grad}. 
    Here, once can see that extensive properties are more relevant than intensive ones for molecular reconstruction. 
    Moreover, we computed the two-dimensional principal component analysis (PCA) of the latent space of the VAE encoder, showing that the higher the  $A$ value, the more correlated is the property with respect to the PCA.
    }
    \label{grad_map}
\end{figure}

\subsection{Interpretability and performance of our model}

We now analyze the established CCS parameterization by implementing a gradient attribution map, which allows us to evaluate the contribution of each property to the output structures.
In doing so, we compute the gradient of the Coulomb matrix components $CM^{k}_i$ of a given molecule $k$ with respect to a property $p_j$, yielding the Jacobian matrix $J^{k}_{ij} = \dfrac{\partial CM^{k}_i}{\partial p_j}$.
By taking the norm of $J^{k}_{ij}$ over the output dimension of CM and, then, averaging over the subset $\mathbb{B}$ of best reconstructed molecules (150 molecules with RMSD $\leq 0.2$ \r{A}), we obtain an attribution map for each property expressed as:
\begin{equation}\label{grad}
    A_j = \frac{1}{N}\sum_{k\in {\mathbb{B}}} \norm{\dfrac{\partial CM^{k}_i}{\partial p_j}},
\end{equation}
where $N$ is the total number of molecules included in $\mathbb{B}$.
Fig.~\ref{grad_map} shows the values of the attribution map $A$ per property which have been normalized over the maximum value and sorted in decreasing order for the full representation.
Overall, we have found that extensive properties are more informative than the intensive ones for the task of molecular reconstruction. 
This can be explained by the fact that these extensive properties depend on  crucial molecular features that are also considered in a 3D representation like CM, \eg number of atoms, number of electrons (related to the chemical composition) and geometry. 
Moreover, when comparing CMs, even a slight difference of one atom can significantly increase the loss, leading to larger sensitivity of the  model to variations in system size and composition. 
Thus, the $A$ values for the components of the total energy and the molecular polarizability are higher compared with these for the molecular orbital energies and dipole moment; in particular, $E_{XX}$ and $E_{KIN}$ present the largest $A$ values. 
Interestingly, this finding is in agreement with the identification of molecular clusters in the two-dimensional principal component analysis (PCA) of the latent space of the VAE encoder, \ie the higher the  $A$ value, the more correlated is the property with respect to the PCA (see insets in Fig.~\ref{grad_map}).
The fact that a linear method such as PCA shows good organization in terms of the energy-related properties is a non-trivial finding which will be later exploited in the interpolation of transition structures.

\begin{figure}[t!]
    \centering
    \includegraphics[width=0.7\linewidth]{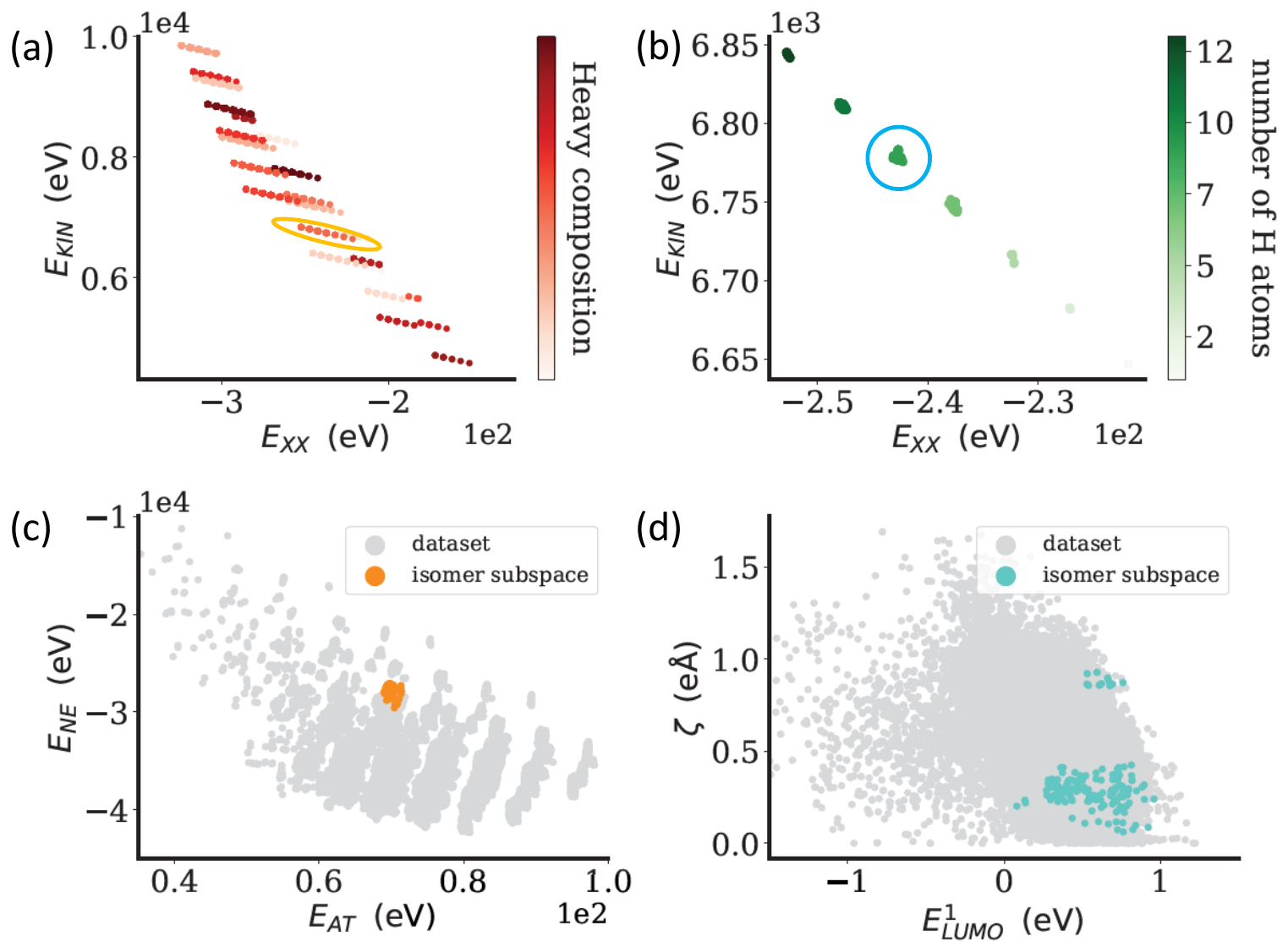}
    \caption{\textbf{Two-dimensional (2D) projections of the high-dimensional QM7-X molecular property space.} 
    We have studied how the revealed hierarchy of QM properties (see Fig.~\ref{grad_map}) organizes the QM7-X chemical compound space.
    (a) 2D property space defined by the two properties with highest $A$ values, \ie $\left(E_{XX},E_{KIN}\right)$-space.  Here, the molecules in the dataset seem to organize as linear-shape clusters containing molecules with the same heavy atom composition (see color code). 
    (b) A closer examination of one of the highly populated clusters in the $\left(E_{XX},E_{KIN}\right)$-space (see yellow ellipse in panel (a)). Molecules are highlighted with respect to their number of H atoms.  
    We then picked a molecular isomer subspace (see blue ellipse in panel (b)) and plot how different pairs of QM properties with (c) high and (d) low $A$ values distribute the molecular structures. 
    This analysis demonstrated that high $A$ values identify properties that are better local coordinates for exploring this subspace, \ie these properties can be used to distinguish molecular structures within a specific molecular isomer subspace.}
    \label{props_analysis}
\end{figure}

Furthermore, we have studied how the revealed hierarchy of QM properties organizes the QM7-X CCS.
Starting from the properties with the highest $A$ value, in Fig.~\ref{props_analysis}(a), one can see the two-dimensional projection of the QM7-X molecular property space defined by $E_{KIN}$ and $E_{XX}$, \ie $\left(E_{KIN}, E_{XX}\right)$-space.
Despite these two properties having a high inverse correlation (Pearson coefficient $=-0.92$), it is noticeable how the molecules in the dataset seem to organize as linear-shape clusters containing molecules with the same heavy atom composition.
In particular, upon examining those clusters,  it becomes evident that $E_{KIN}$ is mostly influenced by the heavy atom composition within a molecule.
On the  other hand, $E_{XX}$ is highly sensitive to the number of H atoms, thereby indicating a dependence on the particular bond types that are present. 
This is further analysed in Fig.~\ref{props_analysis}(b), where we provide a closer examination of one of the highly populated clusters (see yellow ellipse in plot) and highlight the molecules with respect to their number of H atoms. 
%
Here, we uncovered a finer local structure with almost perfect inverse correlation (Pearson coefficient $=-0.99$) as well as very compact clusters formed by isomers. 
Overall, this behavior can be understood by considering the qualitative aspects of $E_{KIN}$  and $E_{XX}$: the dominant contribution to $E_{KIN}$ stems from the inner shell electrons (trivial consequence of the virial theorem) and the primary influence on $E_{XX}$ arises from the valence electrons.
Also, exchange-related quantities have been found to play a significant role in characterizing bonds,~\cite{exchangebond, exchange_bond_2} offering an explanation for their sensitivity to the number of H atoms in a molecule.
%
These considerations account for the efficacy of these two QM properties in identifying clustered molecular isomer subspaces and accurately predicting heavy atom composition.
%

Next, we pick a molecular isomer subspace (see blue ellipse in plot) and show how different pairs of  QM properties with high and low $A$ values distribute the molecular structures (see Figs.~\ref{props_analysis}(c,d)). 
As expected, high $A$ values identify properties that are better local coordinates for exploring this subspace (\textit{vide supra}).
Indeed, these properties present relatively smaller changes in their values across related structures in comparison to properties with low $A$ values---an  example of their efficiency in identifying molecular structures within a specific molecular isomer subspace  while effectively distinguishing them from other structures spanning the entire property spectrum.
%
%
Notice that the same behaviors can be found throughout the entire set of QM7-X equilibrium molecules.  
These findings already provide compelling evidence for the potential of our proof-of-concept implementation in furthering our understanding of the molecular property space and unraveling the intricate relationship between QM properties and molecular structures. 

\begin{figure}[t!]
    \centering
        \includegraphics[width=0.9\linewidth]{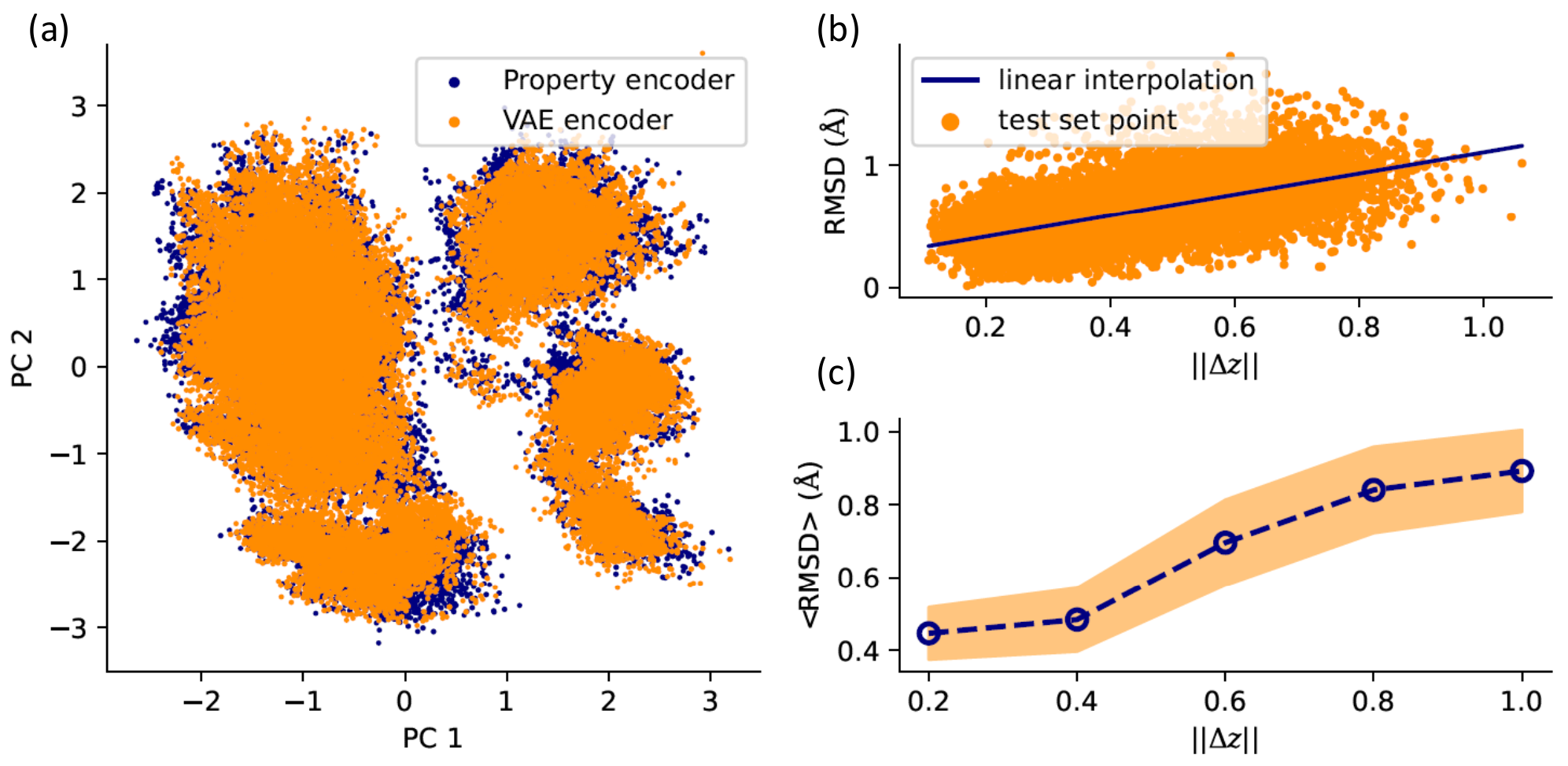}
    \caption{\textbf{Similarity analysis between the latent space representation from the VAE and property encoders.} 
    (a) Overlap between the two-dimensional principal component analysis (PCA) of both latent space representation. 
    (b) Correlation plot of the RMSD between the original and the reconstructed structures versus the latent space difference $||\Delta z|| = ||z-\Tilde{z}||$ for molecules in test set. Here, $z$ is the latent representation from the property encoder, while $\Tilde{z}$ is a new latent representation produced by taking the reconstructed representations and encode them again with the VAE encoder.  
    (c) Average value and standard deviation (orange shadow) of the RMSD corresponding to different values of latent space difference $||\Delta z||$.
    Subsequently, since we are looking for high-quality generated molecules that have low RMSD, we define the interval $||\Delta z|| \in \left[0,0.4\right]$ to filter out molecules during the molecular generation process. }
    \label{latent_diffs}
\end{figure}

The successful property-based parameterization of CCS achieved by the model is grounded in the remarkable similarity attained among the latent representations of CMs and respective properties. 
A visualization that demonstrates this observation can be seen in Fig.~\ref{latent_diffs}(a),  which showcases the  overlap between the PCA of both latent representations.
This verifies the initial assumption we stated in Sec.~\ref{sec:model} about the joint training procedure.
%
%
%
Moreover, we have examined how the differences in latent space representation correlates with the quality of the reconstructed molecules. 
This analysis aims at obtaining a self-consistent method for error estimation.
To this end, if $z$ is the latent representation from the property encoder, we can take the reconstructed representations and encode them again with the VAE encoder, producing a new latent representation $\Tilde{z}$.
Then, we look at the correlation of the quantity $||\Delta z|| = ||z-\Tilde{z}||$ with the RMSD between the original structures and the one reconstructed from properties considering the molecules in test set.
In Fig.~\ref{latent_diffs}(b), one can see that there seem to be a bulk (approximately linear) correlation with the presence of numerous outliers. 
To further investigate this behaviour, we also look at the average value and standard deviation of the RMSD for varying values of $\Delta z$ by splitting the test set into subsets to get an adequate statistics for each point. 
The results are plotted in Fig.~\ref{latent_diffs}(c) in which we show that there is a nonlinear but monotonic behaviour for the relationships between these quantities, finding a minimum for RMSD in the region of low $||\Delta z||$ values ($\in \left[0,0.4\right]$).
Since for reconstructing the actual molecular structure, we are mostly interested in having a low RMSD, we will filter out the generated structures for which $||\Delta z||$ falls outside this predetermined interval in the following sections.
This screening approach based on error estimation will be employed in the next section to enhance the quality of generated structures with a targeted set of QM properties.

\subsection{Novel applications to chemical compound space}

\subsubsection*{Navigating the molecular property space} 

The learned CCS parameterization based on QM properties can provide a versatile solution for multi-objective targeted molecular generation. 
Here, we show the capability of the model to target predefined pairs of QM properties, denoted as $\{\textbf{m}\}$.
In doing so, we first acknowledge that our model has a fixed-dimensional input space for properties, ensuring the attainment of the random sampling associated with typical generative models.
The conditional sampling is then achieved by sampling the non-targeted properties, denoted as $\{\textbf{n}\}$, while conditioning on $\{\textbf{m}\}$ values.
For this purpose, we need a modeled version of the distribution of the property space spanned by QM7-X that facilitates the determination of the posterior distribution of $\{\textbf{n}\}$, given fixed $\{\textbf{m}\}$ values.
Indeed, we use a multivariate Gaussian regression approach to model this property space.
Specifically, we constructed a model with 91 multivariate Gaussian distributions $\{\mathcal{N}(\mu_k, \Sigma_k)\}_{k\in\{1[...]91\}}$ with $\mu_k$ and $\Sigma_k$ as the mean value and the covariance matrix of the Gaussian $k$.
This choice was made following a Bayesian information criterion (see Sec. 3 of SI).

With fixed target values, denoted as $\{\Bar{\textbf{m}}\}$, the Gaussian $\Bar{k}$ from which it is more likely to sample the targeted properties is then selected following the maximum likelihood criterion $\Bar{k} = \underset{k}{\mathrm{argmax}} \left(\mathcal{N}_{\textbf{m}}(\textbf{m}=\Bar{\textbf{m}}|\mu_k,\Sigma_k)\right)$, where $\mathcal{N}_\textbf{m}$ is the marginal distribution relative to the chosen set of properties $\{\textbf{m}\}$. 
The variables with ``\ $\Bar{}$\ '' on top correspond to the target values throughout the text.
Next, the conditional probability formula for multivariate Gaussian distributions is applied, obtaining the distribution of the non-targeted properties $\{\textbf{n}\}$ for $\textbf{m} = \Bar{\textbf{m}}$. 
This can be written as,

\begin{equation}
    p(\textbf{n}|\textbf{m} = \Bar{\textbf{m}}) = \mathcal{N}(\textbf{n}|\Tilde{\mu}, \Tilde{\Sigma}),
\end{equation}

with $\Tilde{\mu}$ and $\Tilde{\Sigma}$ defined as,

\begin{align}
    \Tilde{\textbf{$\mu$}} &= \textbf{$\mu$}_n + \Sigma_{nm}\Sigma_{mm}^{-1}(\Bar{\textbf{m}}-\textbf{$\mu$}_m) \\
    \Tilde{\Sigma} &= \Sigma_{nn} - \Sigma_{nm}\Sigma_{mm}^{-1}\Sigma_{mn}.
\end{align}

To proceed with the generation of molecular structures, multiple values for $\{\textbf{n}\}$ are sampled from $\mathcal{N}(\textbf{n}|\Tilde{\mu}, \Tilde{\Sigma})$ for fixed $\textbf{m} = \Bar{\textbf{m}}$.
Each sample $s$ will correspond to the total set of (targeted and non-targeted) properties  $\Bar{\textbf{m}} \bigoplus\textbf{n}_s$  which is then fed to the model that will generate a Coulomb matrix sample $\textbf{CM}_s$.
At this stage, the VAE encoder is used to encode the samples and, consequently, to obtain the latent representation encoding $z_{s}$.
This is used to filter the samples based on criteria $\Delta z = ||z-z_{s}|| \in \left[0,0.4\right]$ (\textit{vide supra}), where $z$ is the encoding from the property encoder.
Note that this specific interval is adjusted for each target, taking into consideration the possibility of it being either too loose or excessively stringent.
Cartesian coordinates and atomic species of molecules are posteriorly retrieved from the generated Coulomb matrices and OpenBabel is used to add \ch{H} atoms according to the bond orders of heavier atoms.
After optimizing the position of \ch{H} atoms, the generated molecules are screened by analyzing the bond structure and the maximum force component, which must be lower than a threshold (forces were computed with DFTB3+MBD level of  theory). 
This threshold is set to a very high value (10 a.u.) to guarantee that the structure does not contain overlapping atoms without excessively constraining the structure generation. 
%
%
As a final step, the generated structures undergo geometry optimizations using  DFTB3+MBD and, subsequently, their QM properties  are calculated at the PBE0+MBD level of theory to enable a comparison with the target values $\{\Bar{\textbf{m}}\}$ in the QM7-X dataset.

\begin{figure}[t!]
    \centering
        \includegraphics[width=1\linewidth]{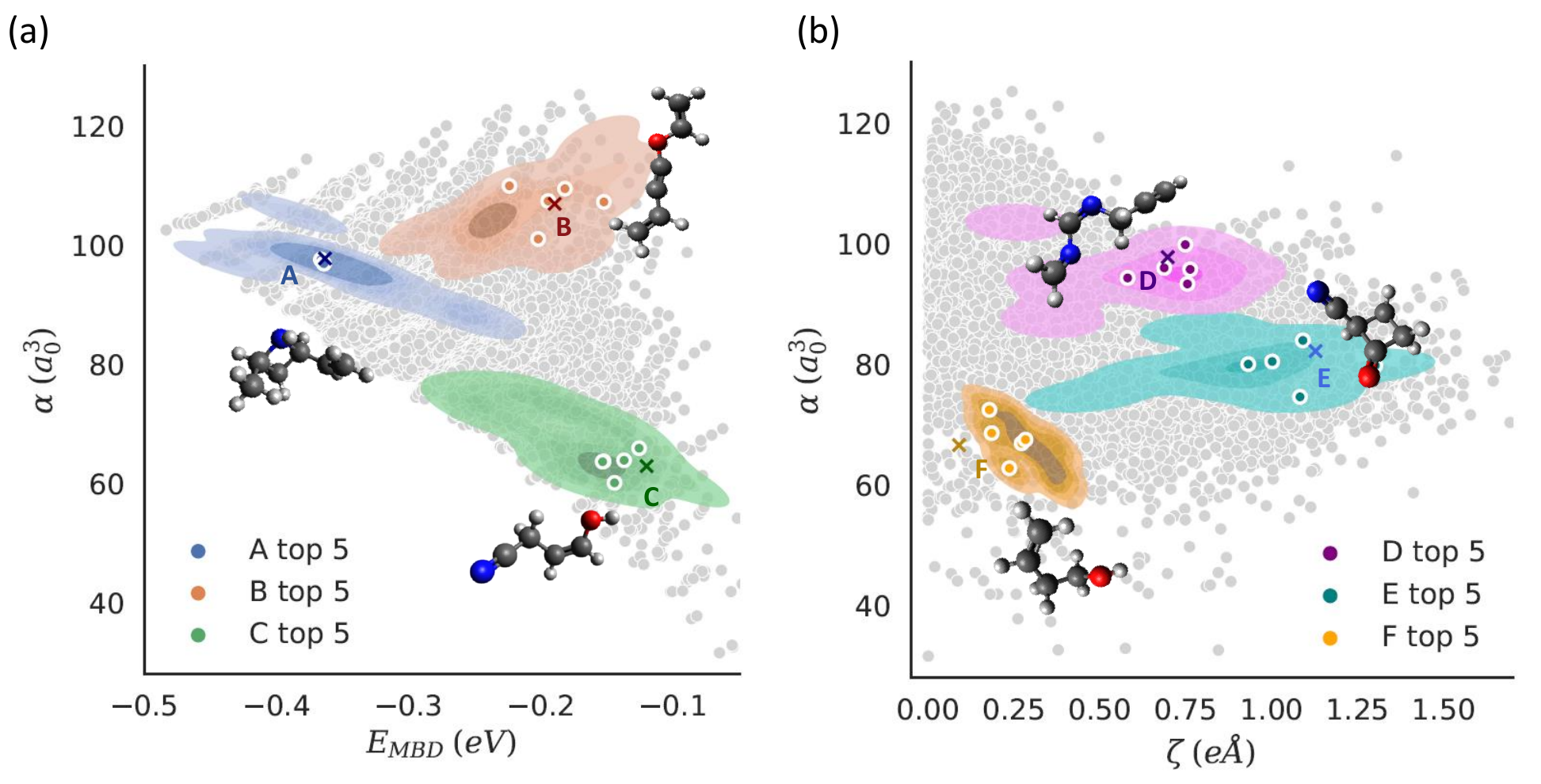}
    \caption{\textbf{Generation of molecules with a targeted pair of quantum-mechanical (QM) properties.} 
Here, we show the distribution of the fifteen generated molecules (or 15-molecule set) per targeted pair of QM properties (colored crosses) in the 2D property space defined by (a) molecular polarizability and MBD energy (\ie $\left( \alpha, E_{MBD}\right)$-space)  and (b) molecular polarizability  and total dipole moment (\ie $\left( \alpha, \zeta \right)$-space).
 To reconstruct these molecules from their QM properties, we have followed the molecular generation procedure explained in  Sec. 3.3. 
The top 5-molecule set (\ie five optimal molecules) per target have been highlighted with colored circles in each panel.
For reference, the values corresponding to QM7-X molecules are shown in the background (grey dots). 
 Panels (a) and (b) contain a single structure per target that represents the top 5-molecule set (other structures are depicted in Fig. S5 of the SI). 
 These results demonstrate that our model displays a  better performance for molecular reconstruction when only targeting   extensive properties (see spread of colored  circles with respect to targeted values), which can be primarily attributed to the purely geometrical/chemical definition of the CM representation.}
 %
 %
    \label{targeted_gen}
\end{figure}

Based on this procedure, we first navigate through the $\left( \alpha, E_{MBD}\right)$-space that encompasses two relevant properties for molecular reconstruction (high attribution value $A$), see Fig.~\ref{targeted_gen}(a).
The moderate degree of correlation between these properties of QM7-X molecules (\ie Pearson coefficient $= 0.60$, grey dots) also grants us access to the intrinsic  ``freedom of design'' in CCS when searching for molecules with desired functionality.~\cite{medranoquantum}
Subsequently, we transition to a weakly correlated 2D property space given by $\left( \alpha, \zeta \right)$-space (\ie Pearson coefficient $= 0.44$, grey  dots) by replacing $E_{MBD}$ with a dipole moment $\zeta$ (low $A$), see Fig.~\ref{targeted_gen}(b).
In both 2D property spaces, fifteen samples (or 15-molecule set) per targeted pair of QM properties (colored crosses labeled  with letters) were generated. 
%
%
The targeted $\alpha$ values were here selected with the aim of generating medium-to large-sized molecules, since $\alpha$ is  an extensive property that mostly depends on the elemental composition  as well as  the connectivity of atoms within the molecule. 
Whereas, the targeted $E_{MBD}$ and $\zeta$  values were chosen to tune the topological effects (\ie extended vs. compact, packed/globular vs. void space) and elemental composition (homogeneous  vs. heterogeneous) in the generated molecules, respectively.  
The top 5 generated molecules (\ie five optimal molecules) per target have been highlighted with colored circles in  both 2D property spaces.
In Fig. ~\ref{targeted_gen},  we also show the generated  structure of a select molecule per target.
Overall, these results demonstrate that our model is capable of generating diverse molecular structures with similar scaffolds as in QM7-X molecules by using only QM properties as input.
Given the discrete nature of CCS and particularly the reduced coverage of the employed QM7-X dataset, it is clear that the inverse mapping generates molecules that can deviate from the respective targeted values.
The performance per target can be quantitatively measured by defining  $\epsilon = \dfrac{|y_{\rm calc} -y_{\rm t}|}{\Delta y}\times100$ with $y_{\rm calc}$ and $y_{\rm t}$ as the calculated and target values of the property $y$.
$\Delta y$ represents the extent of the property spectrum across the entire dataset. 
Accordingly, for each of the 15-molecule set depicted in Fig.~\ref{targeted_gen}(a), the $\epsilon$ value is circa 9.2\% for $E_{MBD}$ and circa 3.5\% for $\alpha$. 
These values are reduced to circa 3.2\% for $E_{MBD}$ and circa 1.3\% for $\alpha$ by only considering the top 5-molecule set.
On the other hand, the molecules generated in $\left( \alpha, \zeta \right)$-space displayed $\epsilon \approx 2.8 \%$ for the prediction of $\alpha$ but the $\epsilon$ corresponding to $\zeta$ was circa $5.9\%$ (see top 5-molecule set per target in Fig.~\ref{targeted_gen}(b)).
%
%

Furthermore, we have found that the spread of the generated molecules (\ie flexibility of the model in molecular generation) can be rationalized by analyzing the relative variance of the conditional multi-Gaussian distributions of the non-targeted QM properties.
In fact, molecules generated with the highest precision (\ie targets A and D) are characterized by lower negative log likelihood values and small relative variances in the extensive non-targeted properties (see Fig.~S4 of the SI).
This outcome could be a key factor for controlling the degree of flexibility when designing molecules in targeted regions of a given property space since a larger variance in extensive properties may result in a more diverse set of molecules as these properties are the most relevant in defining the molecular structure (see Fig.~\ref{grad_map}).
In this regard, taking a closer look at the generated molecules, one can see that the molecules in the $\left(E_{MBD}, \alpha \right)$-space display a greater diversity in heavy atom composition compared to the ones generated in the $\left(\zeta, \alpha \right)$-space, see Fig. S5 of the SI.
Importantly, the  presence of diverse chemical compositions and structures within the 15-molecule set for each target is another evidence that our approach does not follow a conventional chemical exploration based on fixed molecular scaffolds. 
Instead, our inverse design procedure can fully utilize the diversity of property-to-structure relations as embodied in the recently proposed ``freedom of design'' conjecture in CCS.~\cite{medranoquantum}

These findings have thus demonstrated the potential of a CCS parameterization that uses QM properties as intrinsic coordinates to navigate the space of molecular structures spanned by QM7-X dataset.
Unlike the traditional generative paradigm in machine-learned latent spaces, our approach allows to determine and interpret the input conditional distribution since every coordinate in our model is a QM property with a clear physical meaning. 
This is a crucial point, especially when assessing the relation between sample quality and relative variance on extensive properties for the conditional multi-Gaussian distribution.
As a current limitation, we remark that the trained model exhibits certain restrictions in navigating scaffolds that are not present in the dataset, showing a deficiency in handling novel chemistry.
This lack of novelty in molecular generation is mainly related to the systematic design of the QM7-X dataset,~\cite{vignac} as well as the use of a distance matrix-based representation -- both limitations could be relaxed in future work. 
%

\begin{figure}[t!]
    \centering
        \includegraphics[width=0.6\linewidth]{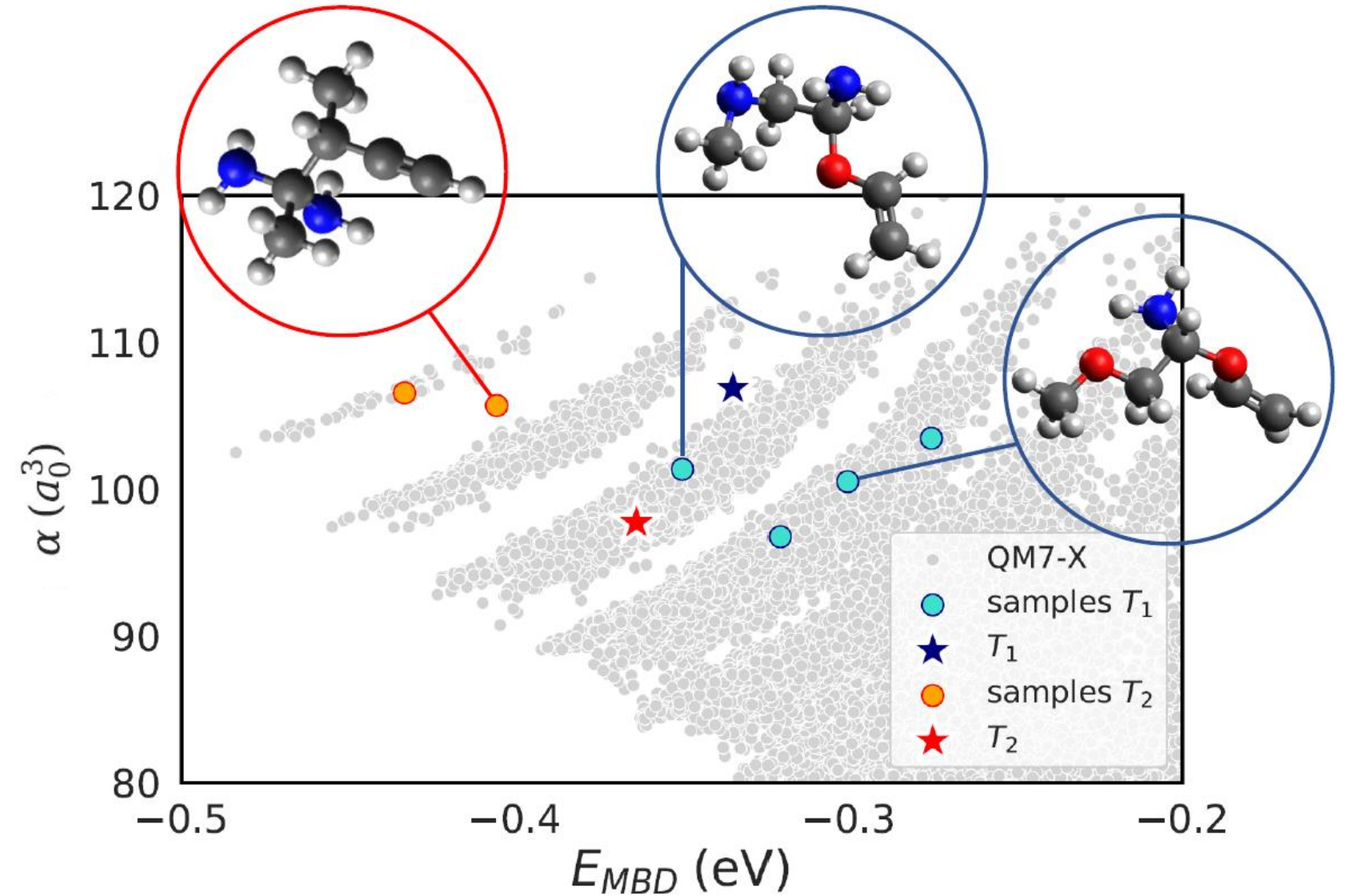}
    \caption{\textbf{Targeting novel compounds with a given pair of molecular properties.} 
    Evaluation of the impact of the modified molecular generation procedure on multiple targets across the $(\alpha,E_{MBD})$-space.
    To do so,  we have retrained the model with a bias towards treating molecular fragments more independently (see more details in Sec. 4 of SI). 
    In the graph, we plot the targets $T_1$ (blue star) and $T_2$ (red star) and their associated molecules with eight heavy atoms composition that were generated using the new model (colored  circles). 
    For reference, the values corresponding to QM7-X molecules are shown in the background (grey dots). 
    Also depicted are select novel molecular structures per target. 
    %
    }
    \label{newchem}
\end{figure}

\subsubsection*{Generating novel compounds in uncharted CCS regions}

In order to demonstrate that our model is capable of generating \textit{de novo} molecules with desired QM properties beyond the QM7-X dataset, we have modified the training procedure to reduce the bias of the model towards treating existing molecular fragments (see Sec. 4 of the SI).
This action slightly reduces the model performance in molecular reconstruction, but it enables the generation of scaffolds beyond QM7-X, mostly composed of novel molecules containing eight heavy atoms.
While using this new procedure across multiple targets in the $\left( \alpha, E_{MBD}\right)$-space, we found that the generation of novel compositions is confined to a region of low coverage defined by high $\alpha$ and large $|E_{MBD}|$ (see Fig. \ref{newchem}). 
Certainly, the model here exhibits a higher degree of flexibility and it is capable of generating molecules with diverse composition of eight heavy atoms. 
In Fig. \ref{newchem}, one can also see that the spread of generated molecules for target $T_{1}$ and $T_{2}$  is broader than that of corresponding targets discussed in the preceding section.
In term of the relative error $\epsilon$, the samples in target $T_{1}$ presented an error circa 10\% for $E_{MBD}$ and 8\% for $\alpha$, while, for samples in $T_{2}$, the observed errors were circa 20\% for $E_{MBD}$ and 16\% for $\alpha$. 
Despite this lower prediction accuracy of QM properties, the errors for target $T_{1}$ are notably comparable to the results shown in Fig.~\ref{targeted_gen}.
Moreover, by analyzing the generated molecules, we found that the novelty in heavy atom compositions can vary depending on the specific target location, \eg $T_{2}$ samples consider only (C,N)-based molecules while $T_{1}$ samples are more chemically diverse and cover (C,N,O)-based molecules.
For both sets of samples, $E_{MBD}$ and $\alpha$ values are in the range of expected values for molecules with eight heavy atoms, \ie $-0.6 \ \mbox{eV} < E_{MBD} < -0.09 \ \mbox{eV}$ and $66 \ \mbox{a.u.} < \alpha < 160 \ \mbox{a.u.}$.
%
%

This initial assessment for using inverse design to generate \textit{de novo} molecules with desired QM properties highlights the promise of rational exploration of CCS, but also uncovers the relative limitation of generative models in extrapolating to molecules larger than the dataset employed for constructing the machine learning model.
Future studies should investigate the use of representations more advanced than the CM matrix and datasets spanning a much larger chemical space than the one given by QM7-X dataset employed in this work.  
%

\begin{figure}[t!]
    \centering
        \includegraphics[width=0.98\linewidth]{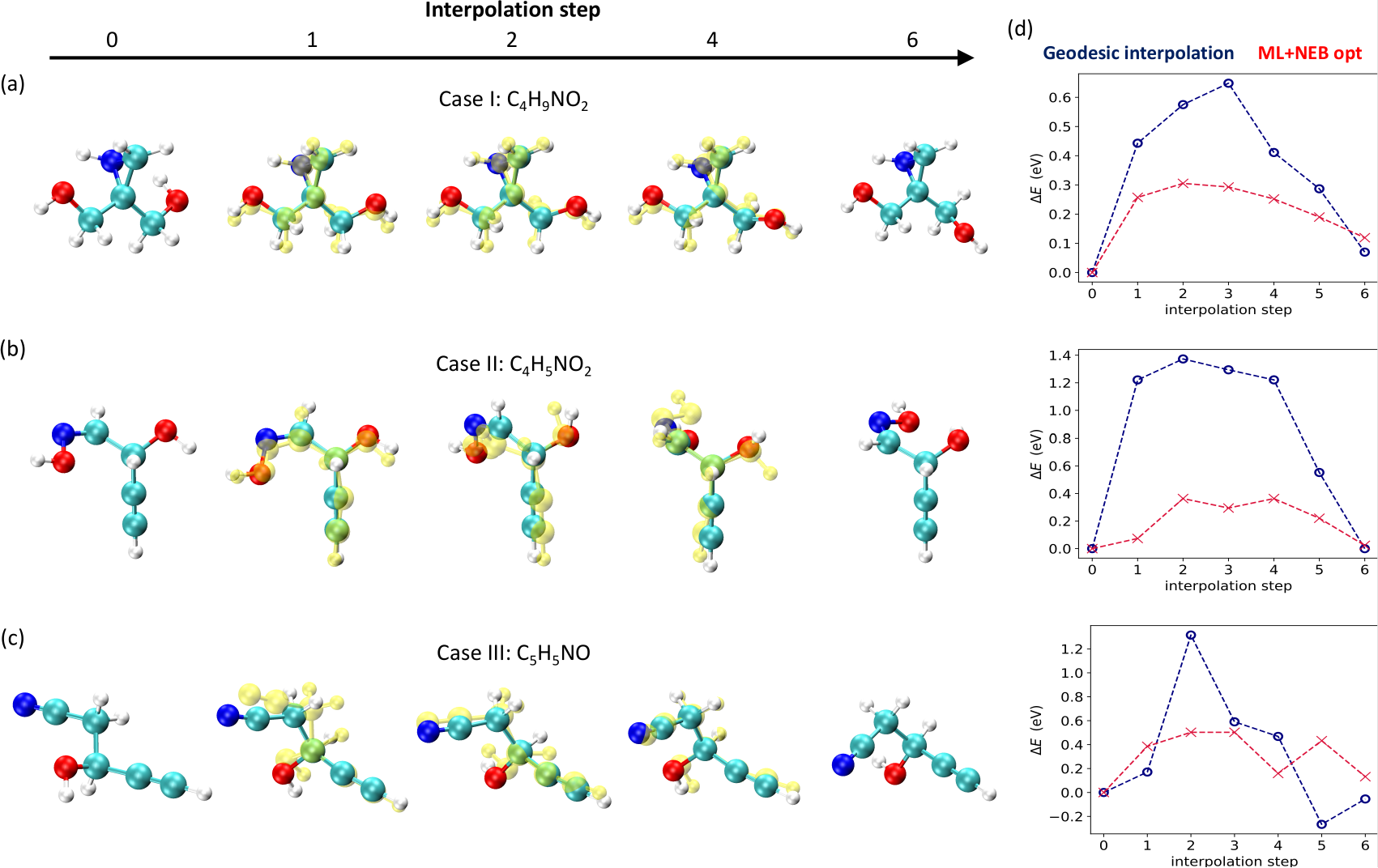}
    \caption{\textbf{Predicting transition path between conformational isomers.}
    %
    %
    By using the geodesic interpolation algorithm for VAEs, we were able to generate interpolated geometries (see yellow balls) for three different pairs of conformational isomers (a) \ch{C4H9NO2} (case I), (b) \ch{C4H5NO2} (case II), and (c) \ch{C5H5NO} (case III).
    For comparison, we show the new transition structures (solid colored balls) obtained by running ML-NEB calculations using as initial guesses the previous interpolated geometries.
    (d) Variation of the relative energy component $\Delta E_{i} = E_{i}- E_{0}$ ($E$ is the sum of atomization and MBD dispersion energies) as a function of the interpolation step $i$ for the three isomerizations shown in panels (a-c).
    We here show the results obtained for the corresponding geodesic in property space (blue curve) and the updated transition structures computed using the ML-NEB method (red curve).
    These findings are compelling evidences that latent representation can be used as an intrinsic coordinate system to generate reasonable guesses for the transition geometries between conformational isomers in QM7-X.}
    \label{interpolate}
\end{figure}

\subsubsection*{Interpolating transition geometries between conformational isomers}

Throughout the preceding sections, we have showcased the versatility of an approximate CCS parameterization in a range of contexts.
Building upon the observation that the latent space of our model exhibits a discernible structure characterized by energetic properties when analysed \via a linear method, we aim to explore the potential of utilizing this latent representation as an intrinsic coordinate system to generate reasonable guesses for the transition geometries between conformational isomers in QM7-X.
While it would be relatively straightforward to perform a linear interpolation in the latent space and generate associated geometries using this model, we are also interested in exploring the feasibility of obtaining an energy profile estimation along with this task.
%
%
To do so, we use the geodesic interpolation algorithm~\cite{geodesic} for VAEs but, in this instance, the objective is to find curves in the property space that are geodesics with respect to the metric induced by the latent space encoding.
This enables us to see how a close-to-linear interpolation in the latent space is reflected in both the property input space and the structure output space. 
The procedure to get the transition geometries starts with the selection of a pair of well-reconstructed conformational isomers (RMSD $\leq 0.2$ \r{A}), namely $\textbf{x}_0$ and $\textbf{x}_N$.
Then, we linearly interpolate $N-1$ points between the corresponding coordinates in property space defined by the seventeen properties considered during the training procedure.
Applying the property-to-structure relationship defined by our model, we obtain an initial configuration given by the set $\{\textbf{y}_0, \textbf{y}_1, \textbf{y}_2, \dots , \textbf{y}_N\}$ of property coordinates, the set $\{\textbf{z}_0, \textbf{z}_1, \textbf{z}_2, \dots , \textbf{z}_N\}$ of latent space coordinates and the set $\{\textbf{x}_0, \textbf{x}_1, \textbf{x}_2, \dots , \textbf{x}_N\}$ of molecular structures.
After this, we use a gradient descent algorithm to minimize the loss function $L = \Sigma_{i = 1}^{N}||\textbf{z}_{i+1}-\textbf{z}_i||^2 + \epsilon\Sigma_{i = 1}^{N}||\textbf{y}_{i+1}-\textbf{y}_i||^2$, where the first term enforces the minimum distance in the latent space, while the second term is a regularization term with $\epsilon<<1$ to enforce continuity in the property space.
To find a stationary path, the optimization runs until a convergence criterion on the norm of the gradient is met (\ie $<10^{-4}$).
Concerning the choice of molecules for this evaluation test, we also took into account that chiral molecules are indistinguishable when using the CM representation. 
When selecting the initial and final states, we ensure that these molecules are not mirror images of one another, as the model would interpret them as identical molecules.
%
%
Consequently, we have selected three different pairs of conformational isomers \ch{C4H9NO2} (case I), \ch{C4H5NO2} (case II), and \ch{C5H5NO} (case III) whose structures are achiral and were reconstructed with an RMSD $\leq 0.2$ \r{A}, see initial ($i=0$) and final ($i=6$) geometries per case in Figs.~\ref{interpolate}(a-c).

Overall, the interpolated geometries displayed in Figs.~\ref{interpolate}(a-c) (see yellow balls) effectively demonstrate the capability of the model to produce plausible geometries for the transition path of the studied isomerizations.
However, by analyzing sample by sample, we detected few abrupt/unphysical changes in the geometries for the cases II and III between the step 1 and 2.
These artifacts primarily arise from two key factors: \textit{i)} the sensitivity of the CM representation to small changes which could produce large mirror-like transformations in the resulting geometry and \textit{ii)} the fact that model performance is degraded across unknown sectors of the latent space associated to the transition geometries. 
The relative energy component $\Delta E_{i} = E_{i}- E_{0}$ ($E$ is the sum of atomization and MBD dispersion energies) of the corresponding geodesic in property space for the three isomerizations is reported in Fig.~\ref{interpolate}(d). 
Unexpectedly, with no request other than minimizing the distance in latent space, a barrier-like behaviour is retrieved in all cases with energy barriers between $0.6 \ \mathrm{eV}$ and $1.4 \ \mathrm{eV}$---another evidence of the potential application of our model for generating guesses of transition geometries as well as energy profiles. 
The obtained results are remarkable since the model was trained exclusively on equilibrium geometries.
%
These findings suggest that the common latent representation of properties and structures is able to capture essential aspects of the underlying physics governing  structure-property relationships, namely, the existence of transition pathways in chemical space. 
%
%

To examine how close the estimated energy profile is to the true minimum energy path and the quality of the generated transition geometries, we have used them as initial guesses for a nudged elastic band (NEB) calculation following the ODE method.~\cite{neb}
Before proceeding with the NEB calculations, \ch{H} atoms were added to all molecules with OpenBabel and, subsequently,  optimizations of initial and final geometries were carried out employing a ML force field trained on PBE0+MBD energies/forces of equilibrium and non-equilibrium molecular conformations contained in QM7-X dataset (model taken from Ref.~\cite{spooky}).
This accurate ML force field was also used to perform the NEB calculations at PBE0+MBD level of theory.~\cite{neuralneb}
The relative energies calculated following this direct procedure are presented in Fig.~\ref{interpolate}(d) and compared to their corresponding geodesic energies, showing how the later were consistently overestimating the energy barrier.
%
%
As for the updated geometries, the RMSD of their heavy atoms structures with respect to the initially interpolated ones was found to be between $0.14$ \r{A} and $0.35$ \r{A}, see colored balls in  Figs.~\ref{interpolate}(a-c).
We remark that the successful execution of a NEB calculation using the generated transition geometries highlights the meaningful nature of the latent representation as intrinsic coordinates in chemical space.
This establishes a connection between our interpolation method and other studies focused on geodesic transition path interpolation.~\cite{geodesic_int}

\section{Conclusions} 
In the present work, we have presented a machine-learning approach to the inverse property-to-structure design process by learning a parameterization of the chemical compound space (CCS) of small organic molecules that uses QM properties as intrinsic coordinates. 
This challenging task was successfully accomplished through the development of a proof-of-concept implementation that jointly trains a variational autoencoder and a property encoder. 
The trained model, utilizing the equilibrium molecules contained in QM7-X dataset, is able to reconstruct the molecules (heavy atom composition and 3D structure) within the test set from their properties with reasonably good accuracy.
Furthermore, the differentiability of the learned CCS parameterization not only enabled us to identify the most relevant properties in the molecular reconstruction process but also revealed intriguing insights. 
Our findings indicate that the combination of exchange energy and kinetic energy plays a pivotal role in clustering the molecules in the dataset based on chemical composition and bond types. 
In contrast, we observed that intensive properties exhibit limitations as local coordinates for conformational space navigation when compared to extensive ones---a result primarily attributed to the purely geometrical/chemical definition of the CM representation.
%

Our trained model has also proven its applicability in diverse molecular design tasks by allowing the conditional sampling for multi-objective molecular generation and the interpolation of transition pathways of chemical transformations. 
Indeed, we have demonstrated that aiming to parameterize the CCS through the utilization of QM properties can lead to the development of more interpretable models capable of performing a wide range of chemical tasks. 
However, our approach still has certain shortcomings that require further investigation in order to obtain a more robust and generalizable model suitable for ML-based screening pipelines within the field of drug discovery.
One thus needs to explore and refine our molecular representation to effectively capture both geometric and electronic features. 
Additionally, it is crucial to expand the studied QM dataset to cover a broader range of chemical and conformational diversity, spanning from small to extended molecules.
We expect this study to serve as a motivation for future research endeavors focused on advancing the field of generative models by leveraging physical and chemical design rules obtained from structure-property/property-property relationships, fostering the development of enhanced models that offer increased accuracy for geometry reconstruction while preserving the high interpretability of the presented approach.

\section*{Acknowledgements} 
The authors acknowledge financial support from the European Union’s Horizon 2020 research and innovation program under the Marie Skłodowska-Curie grant agreement No 956832, “Advanced Machine learning for Innovative Drug Discovery” (AIDD). 
This research used computational resources provided by the High Performance Computing (HPC) of University of Luxembourg.

\section*{Code availability}
The code corresponding to the architecture implementation, training and testing and a notebook to reproduce the main results will be made available on \href{https://github.com/AleFalla/Properties-to-molecules-Inverse-Mapping}{https://github.com/AleFalla/Properties-to-molecules-Inverse-Mapping}.

\bibliography{sample}

\section*{Author contributions}

The work was initially conceived by AF and LMS, and designed with contributions from AT.
AF developed the machine learning code, performed the model training, and analyzed the model performance in diverse  applications together with LMS.
AT supervised and revised all stages of the work. 
All authors discussed the results and contributed to the final manuscript.

\section*{Competing interests}

The authors declare no competing interests.

\end{document}